\documentclass{osa-article}

\journal{osajournal} 
\usepackage{braket}
\usepackage{ mathrsfs }
\usepackage{xcolor,cancel}
\usepackage{multirow}

\newcommand\hcancel[2][black]{\setbox0=\hbox{$#2$}%
\rlap{\raisebox{.45\ht0}{\textcolor{#1}{\rule{\wd0}{1pt}}}}#2} 

\articletype{Research Article}

\begin{document}

\title{Ghost Image Processing}

\author{Harry Penketh\authormark{*}, William L. Barnes, and Jacopo Bertolotti}

\address{Department of Physics and Astronomy, University of Exeter, Exeter, Devon, EX4 4QL, UK
}

\email{\authormark{*}H.Penketh2@exeter.ac.uk} 



\begin{abstract}
In computational ghost imaging the object is illuminated with a sequence of known patterns, and the scattered light is collected using a detector that has no spatial resolution. Using those patterns and the total intensity measurement from the detector, one can reconstruct the desired image. Here we study how the reconstructed image is modified if the patterns used for the reconstruction are not the same as the illumination patterns, and show that one can choose how to illuminate the object,
such that the reconstruction process behaves like a spatial filtering operation on the image. The ability to measure directly a processed image, allows one to bypass the post-processing steps, and thus avoid any noise amplification they imply. As a simple example we show the case of an edge-detection filter.
\end{abstract}


\section{Introduction}

Ghost imaging relies on the combination of two signals which individually are insufficient for image formation \cite{Pittman1995,Bennink2002,Gatti2004}: the sequence of patterns illuminating the object, and the transmitted (or scattered) light, measured with a single element (bucket) detector \cite{Shapiro2008,Bromberg2009,Erkmen2012,Valencia2005, Ferri2005, Zerom2012, Padgett2017,Shapiro2012}.
\\ \indent
In computational ghost imaging one has a great deal of control over the choice of the projected patterns, allowing one to tailor them based on a knowledge of the nature of the object. In Principle Component Analysis the illuminating wavefront is designed to match the principle components of the object \cite{Neifeld2003,Liang2015}, and in other adaptive imaging works the spatial resolution of the wavefront is enhanced locally in response to the detection of high-frequency regions of the object \cite{Assmann2013,Phillips2017}. It is however less common to see the illumination basis modified in response to the way in which the image is to be processed \cite{DelHougne2020}.
\\ \indent
In this paper we show that any post-processing step which can be described by a matrix multiplication with the image, such as convolution with an image filter, can be incorporated into the illumination basis. Doing so enables one to avoid image noise amplification by the filtering process, at the cost of increasing the complexity of the projected patterns. We demonstrate this technique experimentally for a basic edge-detection filter in a modified raster basis, and compare the resulting signal-to-noise ratios (SNRs) with those obtained via post-processing with the same filter. We also discuss a theoretical method to 
predict the performance of an arbitrary filter.

\section{Changing the Illumination Basis}
\begin{figure}[h!]
    \centering
    \includegraphics[width=0.9\textwidth]{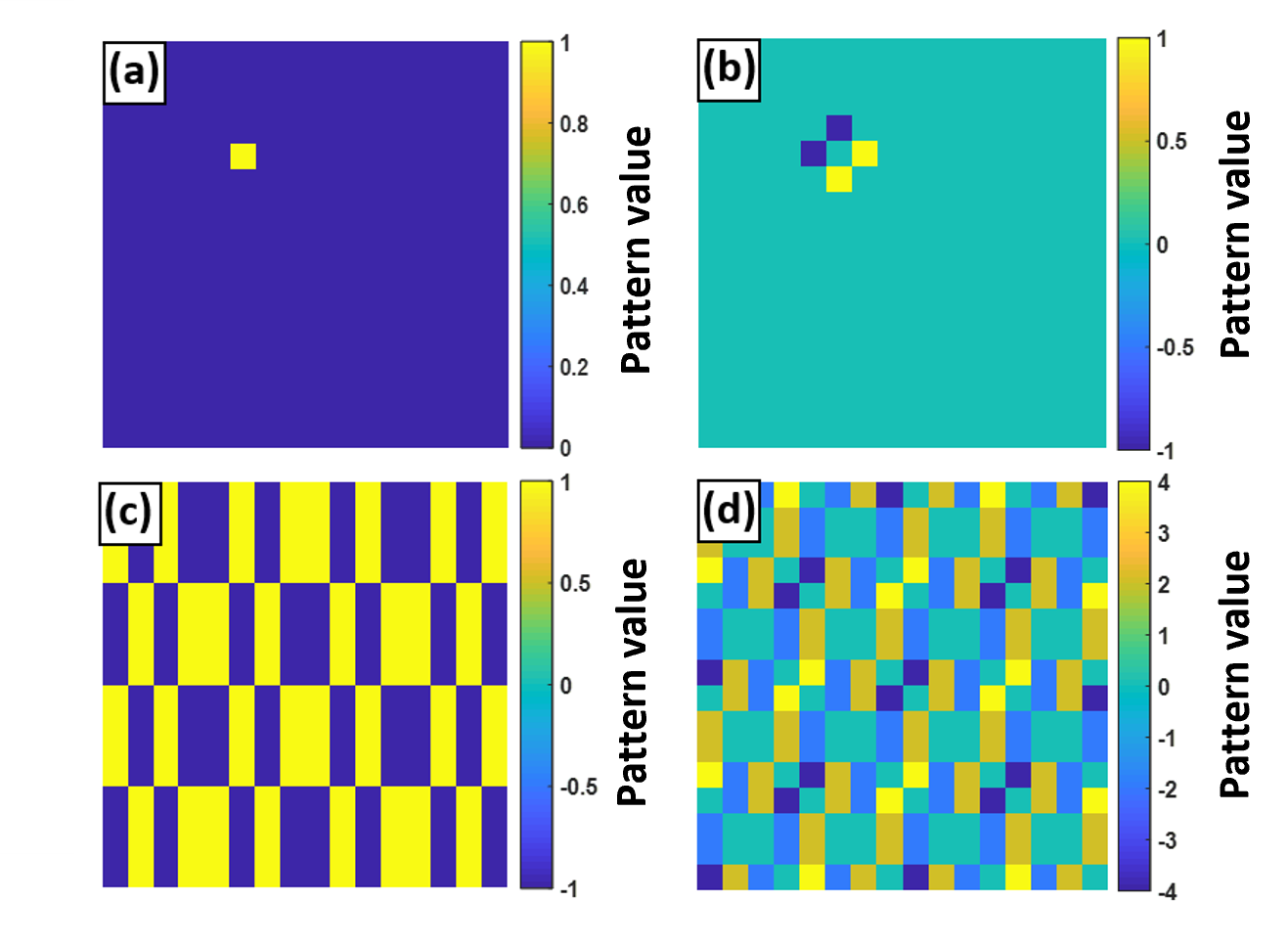}  
    \caption{Examples of illumination patterns before and after multiplication with $B$, in the canonical basis (a) then (b) and Hadamard basis (c) then (d) respectively. The matrix $B$ is such that it performs the edge-detection operation of eq.\ref{eq:K}, with cyclic boundary conditions. The patterns shown are the 85$^{\text{th}}$ in 16$\times $16 resolution bases.}
    \label{FIG:EdgePatterns}
\end{figure}

The ghost imaging measurement process can be described as follows in Bra-Ket notation, with $N \times N$ pixel projection patterns or images represented as $N^2 \times 1$ element column vectors for convenience. The reconstructed image $\ket{I}$ of the object $\ket{O}$ can be written
\begin{equation}
\ket{I} = \sum_j \braket{\psi_j|O}\ket{\psi_j},
\label{eq:GI2}
\end{equation}
where $\ket{\psi_j}$ is the $j$\textsuperscript{th} pattern in the basis $\Psi$ illuminating the object. The inner product $\braket{\psi_j|O}$, which becomes the weighting coefficient of $\ket{\psi_j}$ in the reconstruction, measures the spatial overlap between the projected pattern and object and is recorded with a bucket detector. 
\\
\indent
Typically the illumination basis $\Psi$ is the same basis in which the image is reconstructed, but this need not be the case. A change from an illumination basis $\Psi$ to a new basis $\Phi$ can be written as $\Phi = B \, \Psi$, where B is the matrix that performs the basis change. If one makes this substitution for the illuminating basis in eq.\ref{eq:GI2}
\begin{equation}
\ket{I} = \sum_j \braket{\phi_j|O}\ket{\psi_j} = \sum_j \braket{\psi_j|B^T|O}\ket{\psi_j} = \sum_j \left\langle \psi_j\right\rvert \left( B^T \left\lvert O \right\rangle \right)\ket{\psi_j},
\end{equation}
one can see by comparison with eq.\ref{eq:GI2} that the reconstructed object effectively becomes $(B^T\ket{O})$. The matrix $B$ can then be chosen such that it performs any desired operation, as long as it can be expressed as a matrix multiplication, directly during the measurement process. 
\\ \indent
To demonstrate this equivalence, we chose as our operation convolution with the edge-detection filter kernel $K$: 
\begin{equation}
    K = 
\begin{bmatrix}
0&-1&0\\
-1 & 0 & 1\\
0 & 1 & 0
\end{bmatrix}.
\label{eq:K}
\end{equation}
A 1D convolution between two discrete signals can be written in the form of a matrix multiplication by converting one signal to the appropriate circulant (a subclass of Toeplitz) matrix \cite{Gray2006}. The matrix $B$ that performs the convolution with a given kernel $K$ can be constructed by extending this process to 2D signals with additional zero-padding and flattening steps. Examples of the resulting illumination patterns ($\bra{\psi_j}B^T$), are shown in Fig.\ref{FIG:EdgePatterns} for the cases where the initial basis $\Psi$ is either the canonical or Hadamard basis.

The modified illumination patterns are the result of convolving the original patterns with the filter kernel $K$, providing a more intuitive method of generating the matrix $B$. As with any image convolution a choice of boundary conditions is required, for which we have chosen cyclic conditions which wrap the values at opposite edges \cite{Burger2008}.
\\  \indent 
In the following sections we demonstrate this method experimentally for the edge-detection kernel (eq.\ref{eq:K}) in the canonical basis and compare with the results obtained when the filtering operation is instead applied after reconstruction. 

\section{\textbf{Method}}
\subsection{\textbf{Experimental Setup}}
\begin{figure}[h!]
    \centering
    \includegraphics[width =1\textwidth]{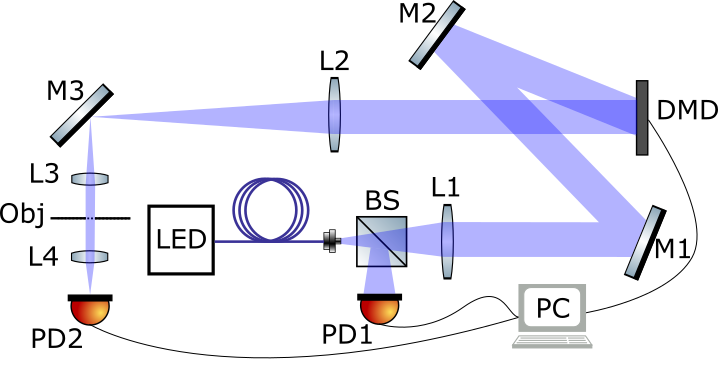}
    \caption{Schematic of the experimental setup. A fibre coupled blue LED is collimated and illuminates a DMD. The DMD is imaged at reduced magnification by lenses L2 and L3 onto a planar transmissive object (Obj). The transmitted light is then collected by lens L4 and focused onto photodiode 2 (PD2). The photodiode 1 (PD1) and beam-splitter combination allow for compensation for fluctuations in the light source intensity.}
    \label{FIG:GISetupDetailed}
\end{figure}

The experimental setup used to compare post-processed ghost images with those generated with a modified illumination basis is shown in Fig.\ref{FIG:GISetupDetailed}. A 455 nm fibre-coupled LED is collimated by plano-convex lens L1 (f = 35 mm) and illuminates the 1080p resolution digital micromirror device (DMD). A beamsplitter allows the LED output to be monitored by a photodiode (PD1). The patterned light from the DMD is then imaged onto the object at reduced magnification by plano-convex lens L2 (f = 200 mm) and biconvex lens L3 (f = 35 mm).
Finally, the light transmitted by the object is collected by biconvex lens L4 (f = 25.4 mm) and focused onto the photodiode PD2. The signal from PD2 can then be divided by the signal from PD1, compensating for fluctuations in the LED output.

\subsection{\textbf{Performing a Fair Comparison}}
One important consequence of modifying the illumination basis is that it might increase the number of unique intensity values required to generate the new projection patterns. This can be seen in Fig.\ref{FIG:EdgePatterns}. To experimentally generate such patterns with a digital micromirror device, which is only capable of binary amplitude modulation, one needs to project and measure multiple times per desired pattern. The number of projectable sub-patterns needed is such that the desired pattern can be formed from a linear combination of sub-patterns. This results in a factor of 2 increase in the total projected patterns when comparing the canonical basis with its edge-detection counterpart (e.g. Fig.\ref{FIG:EdgePatterns}(a) and (b) respectively). We compensate for this in our comparison by repeating each raster pattern twice and using their mean, keeping the total number of projections for each method consistent. Note that the normalisation measurement from the secondary photodiode is performed once per pair of measurements.

\subsection{\textbf{Quantifying the Signal-to-Noise Ratio}}
\begin{figure}[h!]
    \centering
    \includegraphics[width =1\textwidth]{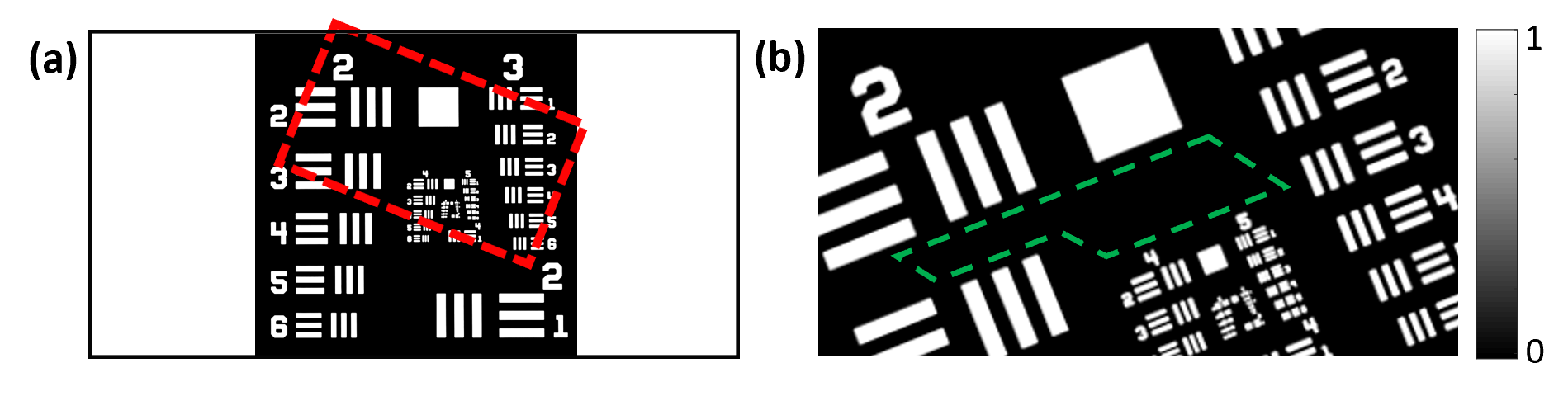}
    \caption{(a) USAF 1951 negative resolution target with the region imaged marked by a red dashed rectangle. (b) The region highlighted in (a), with the green dashed shape indicating the region defined as the background for later SNR calculations.}
    \label{FIG:GhostObj}
\end{figure}

We characterize the quality of the reconstructed images from each method and over a range of detector integration times by means of the signal-to-noise ratio (SNR).
We calculate the image signal-to-noise ratios by comparing the intensity values in defined peak signal ($\langle I_P \rangle$) and background ($\langle I_{B} \rangle$) regions as
\begin{equation}
    \text{SNR} = \frac{\langle I_P \rangle -  \langle I_{B} \rangle}{\sigma_{B}},
    \label{eq:SNRdef}
\end{equation}
where $\langle.\rangle$ denotes the spatial average over a region and $\sigma_B$ the standard deviation of the intensity in the background region. The background region is selected manually and is shown in green in Fig.\ref{FIG:GhostObj}(b). We define the peak signal locations as those with the top 10$\%$ of intensity values, taken from a high SNR experimental image whilst excluding values at the borders. 

\section{Experimental Results}
\begin{figure}[h!]
    \centering
    \includegraphics[width =1\textwidth]{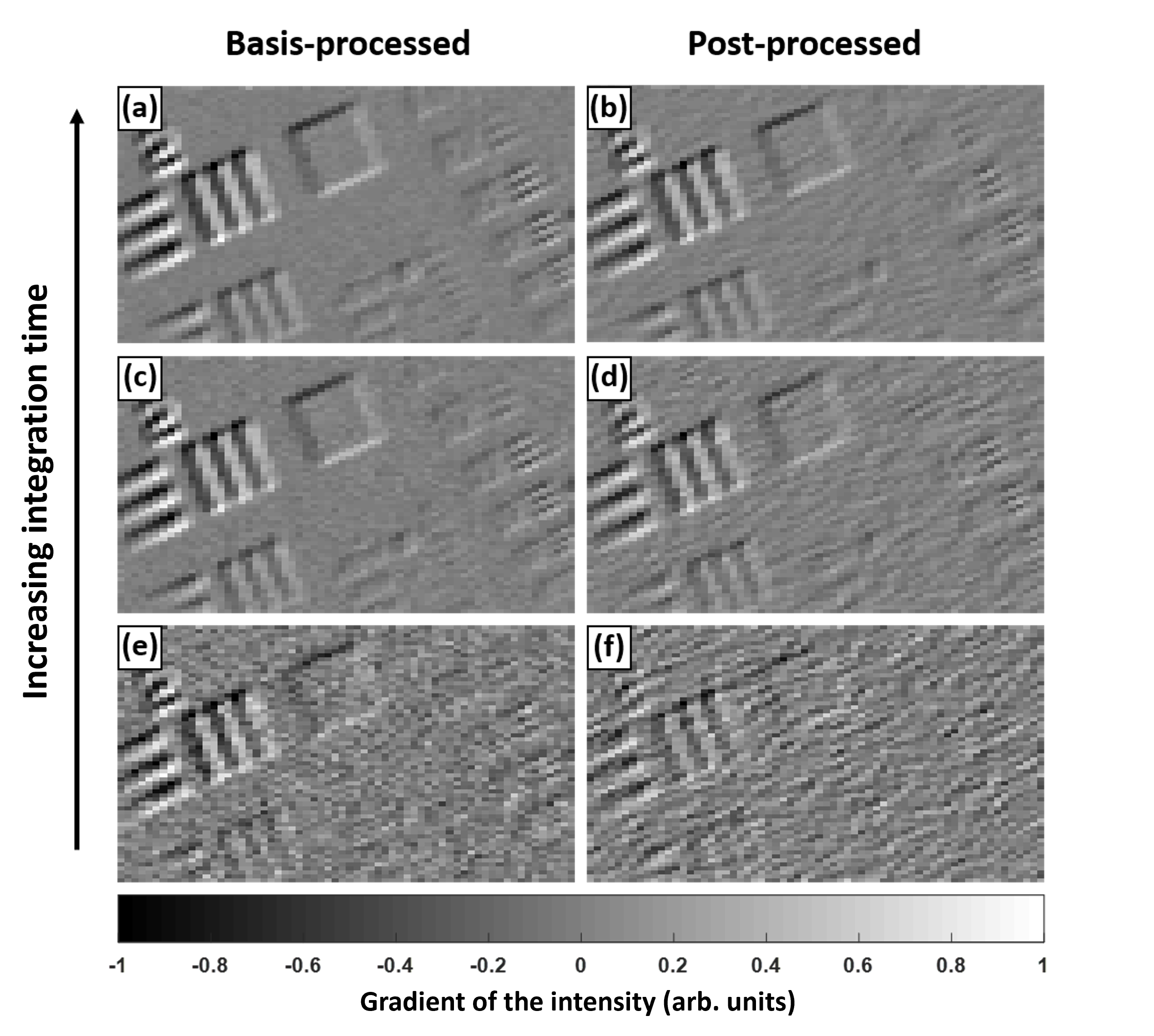}
    \caption{Comparison between the ghost images obtained using a modified projection basis (`basis-processed', left column) and those measured with a raster basis and then convolved with the edge detection kernel $K$ (`post-processed', right column). The three rows show varying detector integration times, increasing from bottom to top as 20, 100 and 220 ms. The images are 64 $\times$ 64 resolution.}
    \label{FIG:GhostVarSNR}
\end{figure}

\begin{figure}[h!]
    \centering
       \includegraphics[width =0.95\textwidth]{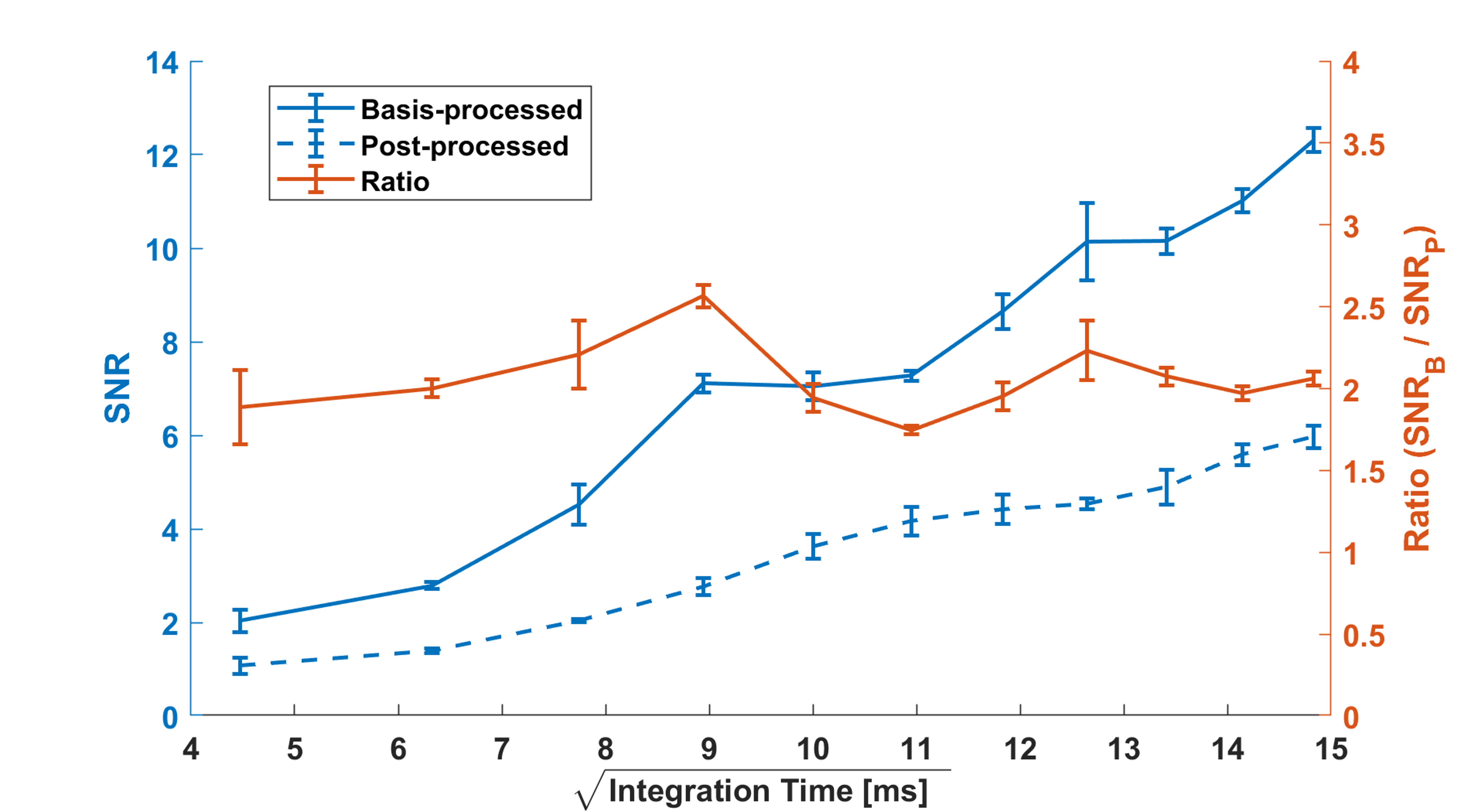}
    \caption{Comparison of calculated SNRs from experimental images acquired using the basis vs post-processing methods.  The error bars are calculated from the variation in three repeat measurements. Images were 64 $\times$ 64 resolution. Lines are a guide to the eye.}
    \label{FIG:GI_SNR_Exp}
\end{figure}

The object, a USAF target as shown in Fig.\ref{FIG:GhostObj}, was imaged in the experimental configuration shown in Fig.\ref{FIG:GISetupDetailed} for a range of integration times. In Fig.\ref{FIG:GhostVarSNR} we show the main result of this paper, a comparison between the images obtained when projecting the modified illumination basis (`basis-processed', left column) and when using a raster basis which is then post-processed by convolution with $K$ (eq.\ref{eq:K}) (`post-processed', right column).
The experimental comparison between post-processing raster images and using a modified illumination basis, shown in Fig.\ref{FIG:GhostVarSNR}, clearly demonstrates that the modified patterns perform the desired spatial filtering operation. When the integration times and thus SNRs of the images are high (Fig.\ref{FIG:GhostVarSNR}(a) and (b)), the difference in image quality is subtle, whilst in the low SNR regime (Fig.\ref{FIG:GhostVarSNR}(e) and (f)) the improvement offered by using a modified illumination basis becomes quite clear.
\\ \indent
The visual differences apparent in Fig.\ref{FIG:GhostVarSNR} have two contributing factors. First, the SNR is higher in the basis-processed case. This is quantified in Fig.\ref{FIG:GI_SNR_Exp} as a factor of 2 enhancement over a range of integration times. The second difference is that the spatial character of the noise has been detrimentally modified in the post-processing case. A post-processing filter creates correlated noise due to the convolution theorem, whilst in the basis-processing approach the noise remains uncorrelated (white) and is less likely to be misinterpreted as part of the object signal.

\section{\textbf{Noise Amplification Model}}
\label{SEC:noiseBasic}
Finally we consider a simple theoretical framework for predicting the difference in signal-to-noise ratios for the two methods and for any convolution kernel $K$. We use additive white Gaussian noise as a model for the detector noise in our measurement system, which we assume dominates. If an image is corrupted by additive white noise, a given filter kernel $K$ will increase the standard deviation of the noise by a factor of $\sqrt{E_k}$, where $E_K$ is the energy of the filter defined as \cite{Jacob2004,Strickland1985}:
\begin{equation}
    E_K = \int_{\Re^2} |K(x,y)|^2\mathrm{d}x\mathrm{d}y.
    \label{eq:FiltNoise}
\end{equation}
For $K$ equal to the edge-detection filter of eq.\ref{eq:K} used in the presented experiments, $E_K$ = 4 and the filter increases the standard deviation of the noise by a factor of 2. The additive Gaussian detector noise introduced with each of the two-step measurements of the object $O$ is denoted $\sigma_1$ and $\sigma_2$ of equal statistics.
 \\ \indent 
 We represent the power normalisation step with a factor $A$, which tracks gradual changes in the lamp intensity with time. The normalisation will generally be imperfect due to two problems: 
first, the measurement of $A$ is noisy in itself, and introduces a new detector noise $\sigma_3$. Second, that either of the photodiode measurements may contain a background, e.g. due to stray light. These are denoted $B_m$ and $B_n$ for the measurement and normalisation signals respectively.
\\ \indent
The signal for the post-processing method $S_P$, comprised of two repeat measurements can be written 
\begin{equation}
\begin{aligned}
     S_P =& \, K * \left[\frac{1}{2}\left(\frac{A O +B_m + \sigma_1}{A +B_n+\sigma_3} + \frac{A O +B_m + \sigma_2}{A +B_n+\sigma_3}\right)\right] \\
     =& \, K * \left[\frac{2AO +2 B_m + \sqrt{2}\sigma}{2A +2 B_n +2\sigma_3} \right] 
     = \, K * \left[\frac{O +B_m/A + \frac{\sigma}{\sqrt{2}{A}}}{1 +B_n/A+\frac{\sigma_3}{A}} \right] =  K * \left[\frac{O +\frac{1}{A} \left( B_m+\frac{\sigma}{\sqrt{2}} \right)}
     {1 +\frac{1}{A} \left(B_n+\sigma_3 \right)} \right] , 
\end{aligned}
\end{equation}
 where $*$ denotes the 2D convolution and independent noise ($\sigma$) terms can be combined in quadrature. In the case where $A$ is large compared to both the noise and the background (i.e. there is a decent amount of signal compared with the artefacts):
\begin{equation}
    \begin{aligned}
    S_P &\simeq K * \left[ \left(O +\frac{1}{A} \left( B_m+\frac{\sigma}{\sqrt{2}} \right) \right) - \left( O +\frac{1}{A} \left( B_m+\frac{\sigma}{\sqrt{2}} \right) \right) \left( \frac{1}{A} \left(B_n+\sigma_3 \right)  \right) \right] \\
    &= K * \left[ O \left( 1 - \frac{1}{A} \left(B_n+\sigma_3 \right) \right)  +\frac{1}{A} \left( B_m+\frac{\sigma}{\sqrt{2}} \right) -\frac{1}{A^2}\left( B_m+\frac{\sigma}{\sqrt{2}} \right)\left(B_n+\sigma_3 \right) \right] \\
   &\simeq  K * \left[ O \left( 1 - \frac{1}{A} \left(B_n+\sigma_3 \right) \right)  +\frac{1}{A} \left( B_m+\frac{\sigma}{\sqrt{2}} \right) -\frac{1}{A^2} \left( B_m B_n + B_m \sigma_3 + B_n \frac{\sigma}{\sqrt{2}} \right)   \right].
    \end{aligned}
\end{equation}
In the final step we apply the filter kernel $K$, denoting the edge-processed version of the object $O_E$ and neglecting the spatial dependence of the normalisation factor $A$, as it varies slowly compared to the size of the filter kernel.

\begin{equation}
    S_P~\approx  O_E \left( 1 - \frac{B_n}{A}\right) - \frac{2 O \sigma_3}{A}   +\frac{\sqrt{2} \sigma}{A} -\frac{2 B_m  \sigma_3 + B_n \sqrt{2} \sigma}{A^2} ,
    \label{eq:PostSignal}
\end{equation}
where the last term is negligible if both $B_m$ and $B_n$ are independent from $A$.
For the `basis-processed' signal $S_B$ we can proceed similarly
\begin{equation}
    \begin{aligned}
    S_B &= \frac{(K_1 * [AO+B_m]) + \sigma_1 - (K_2 * [AO+B_m]) + \sigma_2}{A + B_n + \sigma_3} \\
    &= \frac{K * [O+ \frac{B_m}{A}] + \frac{\sqrt{2} \sigma}{A}}{1+ \frac{B_n + \sigma_3}{A} }
    \simeq \left( O_E + \frac{\sqrt{2} \sigma}{A} \right) \left( 1- \frac{B_n + \sigma_3}{A} \right) \\
   &= O_E \left( 1- \frac{B_n + \sigma_3}{A} \right) + \frac{\sqrt{2} \sigma}{A} - \frac{B_n \sqrt{2} \sigma}{A^2} ,
    \end{aligned}
    \label{eq:basisprocessedSNR}
\end{equation}
where $K * B_m$ is zero if $B_m$ is flat and the last term is negligible if $B_n$ is independent from $A$. \\ \indent 
Any comparison between eq.\ref{eq:PostSignal} and eq.\ref{eq:basisprocessedSNR} requires one to make some assumptions about which term will be negligible and which will dominate. If the signal is much stronger than any stray light or noise, we predictably get $S_B=S_P$. If we keep the terms divided by $A$ (but neglect those divided by $A^2$), we see that $S_P$ is twice as susceptible to the effect of $\sigma_3$ than $S_B$. But in general the exact ratio between the two SNRs is very dependent on the exact parameters of the experiment, and we expect different outcomes depending on the specific filter kernel chosen.

\section{Conclusions}
At its very core ghost imaging is the choice of basis patterns (the illuminations), the measurement of the coefficients (the intensity measured by the bucket detector), and the reconstruction of the image using the two. A big advantage of the ghost imaging method is the total freedom in the choice of the basis used, and we have shown that the freedom of measuring the coefficient of the expansion in a different basis from the one used for the reconstruction allows for even more freedom. In particular one can use this freedom to recover any linear map of the image, thus skipping the post-processing step. We have shown this experimentally for the simple example of an edge-detection filter.\\
\indent
We also studied how the signal-to-noise ratios compare between performing a post-processing filtering, and directly measuring the processed image using ghost imaging. Whether the ghost imaging approach results in a smaller amount of noise depends on the linear map/filter used, and on the experimental details (e.g. the noise in the intensity normalization), but in our experiment we found that the ghost imaging SNR was a factor 2 better than the equivalent post-processed image.

\section*{Funding}
This work was supported by the Leverhulme Trust’s Philip Leverhulme Prize. We acknowledge support from the Engineering and Physical Sciences Research Council (EPSRC) of the United Kingdom, via the EPSRC Centre for Doctoral Training in Metamaterials (Grant No. EP/L015331/1).

\section*{Acknowledgements}
The authors acknowledge useful discussions with David B Phillips.

\section*{Disclosures}
The authors declare no conflicts of interest.

\section*{Data Availability} 
 Data created during this research are openly available from \cite{zenodoDataGI}

\bibliography{sample}
\end{document}